\documentclass[twocolumn,prl]{revtex4}

\begin{document}
\input epsf

\title{Lower bound on the number of Toffoli
gates in a classical reversible circuit through quantum information
concepts}

\author{
Sandu Popescu$^{a,b}$, Berry Groisman$^a$ and Serge Massar$^c$}
\affiliation{
$^a$HH Wills Physics Laboratory,
University of Bristol, Tyndall Avenue, Bristol, BS8 1TL, UK\\
$^b$Hewlett-Packard Laboratories, Stoke Gifford, Bristol
BS12 6QZ, UK\\
$^c$Laboratoire d'Information Quantique and QUIC, Universit{\'e}
Libre de Bruxelles, C.P. 165/59, Boulevard du Triomphe, 1050
Bruxelles, Belgium }

\begin{abstract}
The question of finding a lower bound on the number of Toffoli
gates in a classical reversible circuit is addressed. A method
based on quantum information concepts is proposed. The method
involves solely {\it concepts} from quantum information - there is
no need for an actual physical quantum computer. The method is
illustrated in the example of classical Shannon data compression.
\end{abstract}


\maketitle

In the past ten years we have witnessed the birth and explosive
growth of the field of quantum information and computation. The
main thrust of this new field is to study how quantum systems
(such as quantum computers and quantum communication devices) can
be used to solve certain mathematical problems or to improve
communication capabilities. The crucial feature of this approach
is that although the quantum systems themselves can be studied
with pen and paper, gains are obtained only when the quantum
systems are actually used in practice. The gains are due to new
physical behavior unique to quantum systems and not shared by
classical ones. In this Letter we take a different direction. We
are not interested in using quantum systems, rather we want to use
the concepts and insights gained in the study of quantum
information for solving mathematical problems.

The problem we consider here concerns lower bounds on reversible
classical circuits. Although reversible classical computation will
probably not be realized in the forseeable future (though
increasing attention is being devoted to this issue), its study
has yielded profound insights into the theory of complexity and
into thermodynamics; see \cite{B} for a review.

A reversible classical computation evaluates a function $f$ which
takes $n$-bit input $\bar{x}\in \{0,1\}^n$ to $n$-bit output
$f(\bar{x})\in\{0,1\}^n$. Each particular input has its own unique
output, thus $f$ is invertible. A classical circuit that evaluates
$f$ can be reduced to a sequence of elementary reversible logical
gates. Examples of reversible one-, two- and three-bit gates are
NOT, Controlled-NOT (C-NOT) and Toffoli
(Controlled-Controlled-NOT) gates. C-NOT applies NOT on the second
bit only if the value of a first bit is 1; Toffoli applies NOT on
the third bit only if the values of both first and second bit are
1. Reversible one- and two-bit gates do not constitute a universal
set of gates. The Toffoli gate however is a universal basic gate
for reversible classical computation, i.e., any reversible
classical circuit can be built up from Toffoli gates
\cite{toffoli}.

Although we can build any reversible circuit out of Toffoli gates
alone, an interesting conceptual question is to find the {\it
minimal} number of Toffoli gates required if one allows for {\it
any} number of one- and two-bit gates. The problem is interesting
because Toffoli gates are, in a sense, the strongest reversible
gates, and the minimal number needed tells us about the complexity
of the computation itself. Furthermore, Toffoli gates require {\it
physical interaction} between three bits, and are therefore more
difficult to implement in practice than one- and two-bit gates,
and it might be useful to minimize their use.

We formulate the problem as follows: {\it given a
reversible function $f(\bar{x})$ what is the minimum
number of Toffoli gates needed to construct a circuit
that will evaluate $f(\bar{x})$ for every $\bar{x}$ or only for a
certain subset of $\bar{x}$.}

As far as we know, a systematic approach to this problem does not
exist. In this paper we use quantum information concepts to
address it. In quantum information (computation) one classical bit
can be encoded in two orthogonal states of a quantum system. The
main idea of our method is to {\em map} the bits onto some special
quantum states, and the action of the logic gates onto unitary
transformations acting on these states. Then, the study of the
properties of the unitary transformation that is associated with
the classical reversible computation will give information about
the classical circuit. The map is
\parbox{0.3in}{\begin{eqnarray*}\end{eqnarray*}}\hfill
\parbox{2.in}{\begin{eqnarray*} 0
&\rightarrow& |{\bf
0}\rangle=\frac{1}{\sqrt{2}}(|0\rangle_A|0\rangle_B+|1\rangle_A|1\rangle_B),
\\
1 &\rightarrow& |{\bf
1}\rangle=\frac{1}{\sqrt{2}}(|0\rangle_A|0\rangle_B-|1\rangle_A|1\rangle_B).\label{map}
\end{eqnarray*}}\hfill
\parbox{0.5in}{\begin{eqnarray}\end{eqnarray}}

The states $|{\bf 0}\rangle$ and $|{\bf 1}\rangle$, the ``logical"
qubits into which the classical bits are mapped, represent
entangled states of two ``constituent" qubits, denoted by the
indexes $A$ and $B$. (Throughout this Letter we will use boldfaced
fonts for the logical qubits and normal fonts for the constituent
qubits.) Here the states $|0\rangle,|1\rangle$ are associated with
orthogonal states of a two-level quantum system.

A string of $n$ bits is mapped on the associated quantum state of
$n$ qubit pairs: $ x_1x_2\ldots x_n \rightarrow |{\bf
x}_1\rangle|{\bf x}_2\rangle\ldots|\bf{x}_n\rangle$. Any
computation $x_1x_2\ldots x_n\rightarrow f_1(\overline
x)f_2(\overline x)\ldots f_n(\overline x)$ is mapped on the same
transformation of the corresponding quantum states $|{\bf
x}_1\rangle|{\bf x}_2\rangle\ldots |\bf{x}_n\rangle\rightarrow
|{\bf f_1}(\overline x) \rangle|{\bf f_2}(\overline
x)\rangle\ldots|\bf{f_n}(\overline x)\rangle.$ Since we consider a
reversible classical computation the corresponding quantum
transformation is unitary (or part of a unitary if we we are
interested only in a partial truth table). For example the action
of a C-NOT gate $a,b \rightarrow a,a\oplus b$ ($a,b\in\{0,1\}$) is
mapped on the unitary transformation $U_{C-NOT}$: $|{\bf
a}\rangle|{\bf b}\rangle \rightarrow |{\bf a}\rangle|{\bf a\oplus
b}\rangle$.

The most important property of this mapping is that any classical
reversible circuit built only from one- and two-bit gates is
mapped onto a transformation that requires neither entanglement
nor classical communication, ie. onto
a {\it local} unitary transformation $U=U^A\otimes
U^B$ where $U^A$($U^B$) acts only on the $A$($B$) constituent
q-bits. For
example the $U_{C-NOT}$ gate can be built from {\it local} C-NOT
gates:
\begin{equation}U_{C-NOT}=\tilde U^A_{C-NOT} \otimes
\tilde U^B_{C-NOT}\end{equation} as can be easily checked
explicitly. Here $\tilde U^A_{C-NOT}$ ($\tilde U^B_{C-NOT}$) is a
C-NOT gate acting on the A(B) constituent q-bits in the opposite
direction to normal, i.e. with the second bit as the control and
the first bit as the target. (Such bi-lateral transformations were
considered in \cite{BBPSSW} for the purpose of density matrix
purification). That the quantum equivalent of {\it any} reversible
one or two-bit gates can be constructed by a similar local
bi-lateral transformation can be verified explicitly. Hence, any
circuit built from one or two q-bit gates is local.

We can use this property of our mapping to analyze general
circuits: {\it Given a classical reversible computation we
construct the associated quantum unitary transformation $U$; if $U$
is non-local, the corresponding classical transformation cannot be
constructed solely by two-bit gates. Furthermore, the amount of
non-locality in $U$ gives a lower bound on the number of Toffoli
gates we need.}

We define $E_U$, the amount of non-locality in $U$, as the minimum
amount of entanglement required to implement $U$ using only local
operations and classical communication (LOCC). We denote by $E_T$
the amount of non-locality of the quantum Toffoli gate $U_T$ (we
estimate $E_T$ below).

One possible implementation of $U$ is to
realize the classical circuit using the non-local quantum Toffoli
gates (which cost $E_T$ ebits) and the local 2-bit and 1-bit
gates. Hence $U$ can be implemented using $N_T E_T$ ebits. This
yields the lower bound:
\begin{equation}\label{bound}
N_T\geq E_U / E_T .
\end{equation}

We now arrive at the crucial point of the method. To determine
$E_U$ may be a very complicated task - it might actually be as
complicated as directly determining the required number of Toffoli
gates. On the other hand, it is easier to obtain bounds on $E_U$.
Upper bounds are obtained by explicitly describing an implemention
$U$ and calculating the amount of entanglement required to carry
out this implementation using LOCC. Equation (\ref{bound}) is
derived from just such an upper bound.

To obtain lower
bounds on $E_U$ we apply $U$ on a {\it test state}
$|\Psi_{in}^{test}\rangle$:
\begin{equation}
U|\Psi_{in}^{test}\rangle=|\Psi_{out}^{test}\rangle, \label{tr}
\end{equation}
where the test state can be any arbitrary superposition of basic
input states $|{\bf x}_1\rangle|{\bf
x}_2\rangle\ldots|\bf{x}_n\rangle$. We denote the amount of
non-locality between A and B possessed by $|\Psi_{in}^{test}\rangle$
and $|\Psi_{out}^{test}\rangle$ by $E_{in}^{test}$ and
$E_{out}^{test}$ respectively, where
$E=S(Tr_A|\Psi\rangle\langle\Psi|)=S(Tr_B|\Psi\rangle\langle\Psi|)$
is the von Neumann entropy of the reduced density matrix. (Applying
$U$ to the test state and computing $E_{in}^{test}$ and
$E_{out}^{test}$ is straightforward). The amount of non-locality in
$U$ is not less than the entanglement difference between the two
states:
\begin{equation}
E_U\geq |E_{in}^{test}-E_{out}^{test}|\ .
\label{test}
\end{equation}

How good are these bounds on $N_T$? First of all note that {\it
any} test state leads to a lower bound. However, different test
states may lead to different lower bounds because the non-local
content of $U$ may not be realized in full when $U$ acts on a
particular state. (For example, a test state of the form
$|\Psi_{in}^{test}\rangle=|{\bf x}_1\rangle|{\bf
x}_2\rangle\ldots|\bf{x}_n\rangle$ is transformed into $
|\Psi_{out}^{test}\rangle=|{\bf f}_1\rangle|{\bf
f}_2\rangle\ldots|\bf{f}_n\rangle$ and leads to no increase in
entanglement). Good test states can be found either by trial and
error, or by systematic optimization, although it is unclear
whether the method of test states can provide tight bounds on
$E_U$.

 A more important restriction is due to the fact that
 Eq. (\ref{bound}) can be far from tight. This is because
when implementing the classical circuit some of the Toffoli
 gates may increase the entanglement whereas others
may decrease it. Thus there may be more efficient ways
 of implementing $U$ than realizing the classical circuit.
For instance if $U$ acts on states composed of $n$ logical qubits,
then $E_U\leq 2n$, because one can always implement $U$ by
teleporting Alice's qubits to Bob, letting Bob implement $U$
locally, and teleporting Alice's qubits back to her. This shows
that our method can only provide bounds that grow
 linearly in $n$. On the other hand
it is known that for some problems of classical reversible
computation the number of Toffoli gates grows exponentially
\cite{scott} and for these problems our method
 is very inefficient.
Nevertheless we expect that in many cases the number of Toffoli
gates will grow linearly with, or as a fractional power of, $n$.
Bounding the actual power may give an interesting - indeed,
sometimes fundamental - insight.

Let us now consider the case of the Toffoli gate itself and prove
the basic fact that classical Toffoli gates cannot be built from
reversible two-bit gates. We will do this by showing that under
our map the quantum equivalent of the Toffoli gate is non-local.
Specifically we will obtain the upper and lower bounds
\begin{equation}
1 \leq E_T \leq 2\ . \label{ET}
\end{equation}
Note that it is not essential for our
method to find the exact value of $E_T$ since in general we are
interested only in the scaling of the number of Toffoli gates with
the size of the problem.

The lower bound $E_T\geq 1$ is obtained by showing that under our
map the quantum Toffoli gate is capable of producing at least one
ebit of entanglement. Consider the test state
\begin{eqnarray}
|\Psi_{in}^{test}\rangle={1 \over2}
 (|{\bf 0}\rangle_1|{\bf
0}\rangle_2|{\bf 0}\rangle_3+|{\bf 1}\rangle_1|{\bf
0}\rangle_2|{\bf 0}\rangle_3 +|{\bf 0}\rangle_1|{\bf
1}\rangle_2|{\bf 0}\rangle_3~~~~~\nonumber\\- |{\bf
1}\rangle_1|{\bf 1}\rangle_2|{\bf 1}\rangle_3)
 ={1 \over{\sqrt{2}}}|001\rangle_A|001\rangle_B+{1 \over{2\sqrt{2}}}
\biggl(|000\rangle_A|000\rangle_B
\nonumber\\+|010\rangle_A|010\rangle_B
+|100\rangle_A|100\rangle_B-|110\rangle_A|110 \rangle_B
\biggr),\nonumber \end{eqnarray} where the third logical bit is
the target of the Toffoli gate.  After acting with $U_T$ on
$|\Psi_{in}^{test}\rangle$ we obtain
\begin{eqnarray*}
|\Psi^{test}_{out}\rangle= {1 \over2}(|{\bf 0}\rangle_1|{\bf
0}\rangle_2|{\bf 0}\rangle_3+ |{\bf 1}\rangle_1|{\bf
0}\rangle_2|{\bf 0}\rangle_3
+|{\bf 0}\rangle_1|{\bf 1}\rangle_2|{\bf 0}\rangle_3~~~~~~\nonumber\\
- |{\bf 1}\rangle_1|{\bf 1}\rangle_2|{\bf 0}\rangle_3)={1
\over{2\sqrt{2}}}
\biggl(|000\rangle_A|000\rangle_B+|001\rangle_A|001\rangle_B~~~~~~\\
+|010\rangle_A|010\rangle_B+|100\rangle_A|100\rangle_B
+|011\rangle_A|011\rangle_B~\\
+|101\rangle_A|101\rangle_B-|110\rangle_A|110\rangle_B-|111\rangle_A|111
\rangle_B \biggr).~
\end{eqnarray*} The Schmidt coefficients are found to be $\{\alpha_i\}=\{{1
\over2}, {1 \over 8},{1 \over 8},{1 \over 8},{1 \over 8},0,0,0\}$
and $\{\beta_i\}=\{{1 \over8}, {1 \over 8},{1 \over 8},{1 \over
8},{1 \over 8},{1 \over 8},{1 \over 8},{1 \over 8}\}$
respectively. Hence $E_{in}^{test}=2$ ebits and $E_{out}^{test}=3$
ebits and $E_T\geq 1$. (In passing we note that the quantum
Toffoli gate cannot be implemented without classical
communication: if such an implementation were possible it would
violate relativistic causality).

\begin{figure}
\epsfxsize=3.4truein \centerline{\epsffile{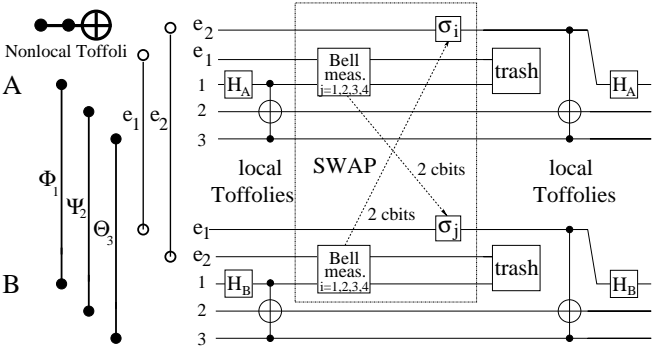}}
\caption[]{Implementation of nonlocal Toffoli using 2 ebits as a
resource.} \label{nonlocalT}
\end{figure}

To obtain the upper bound $E_T \leq 2$ we will describe explicitly
a method for realising the quantum map of the Toffoli that
requires 2 ebits. Consider three pairs of qubits on which we are
going to apply $U_T$, where the states $|{\bf\Phi}\rangle_1$,
$|{\bf \Psi}\rangle_2$ are control and $|{\bf\Theta}\rangle_3$ is
a target (see Fig. \ref{nonlocalT}).
It is convenient to analyze in parallel the cases where
$|{\bf\Phi}\rangle_1$ is  $|{\bf 0}\rangle_1$ or $|{\bf
1}\rangle_1$. The two parties start by performing local Hadamard
rotations $H_{A(B)}$ [acting as $H|0\rangle={1 \over
\sqrt{2}}(|0\rangle+|1\rangle)$, $H|1\rangle={1 \over
\sqrt{2}}(|0\rangle-|1\rangle)$] of $A_1$ and $B_1$ of the first
pair, obtaining

\parbox{0.2in}{\begin{eqnarray*}\end{eqnarray*}}\hfill
\parbox{2.5in}{\begin{eqnarray*}
|{\bf 0}\rangle_1 &\rightarrow& |{\bf 0'}\rangle_1=
\frac{1}{\sqrt{2}}(|0\rangle_1^A|0\rangle_1^B
+|1\rangle_1^A|1\rangle_1^B),\\ |{\bf 1}\rangle_1 &\rightarrow&
|{\bf 1'}\rangle_1 =
\frac{1}{\sqrt{2}}(|0\rangle_1^A|1\rangle_1^B+
|1\rangle_1^A|0\rangle_1^B).
\end{eqnarray*}}\hfill
\parbox{0.3in}{\begin{eqnarray}\end{eqnarray}}

Then the parties proceed by performing local Toffoli gates on
their particles, which can be written as
\begin{equation}\label{localT}
U_{T}^{A(B)}=|0\rangle\!\langle 0|_1 \otimes I_2 \otimes
I_3+|1\rangle\!\langle 1|_1 \otimes U_{23},
\end{equation}
where $U_{23}$ is a local C-NOT between particles 3 and 2 (with
particle 3 as the control and particles 2 as the target). As a
result the initial states evolve to:
\begin{eqnarray}
|{\bf 0'}\rangle_1|{\bf \Psi}\rangle_2 |{\bf \Theta}\rangle_3
\rightarrow \frac{1}{\sqrt{2}}
(|0\rangle_1^A|0\rangle_1^B+|1\rangle_1^A|1\rangle_1^B U_{23}^A
U_{23}^B) |{\bf \Psi}\rangle_2
|{\bf \Theta}\rangle_3,\nonumber\\
|{\bf 1'}\rangle_1|{\bf \Psi}\rangle_2 |{\bf \Theta}\rangle_3
\rightarrow \frac{1}{\sqrt{2}} (|0\rangle_1^A|1\rangle_1^B
U_{23}^B+|1\rangle_1^A|0\rangle_1^B U_{23}^A) |{\bf \Psi}\rangle_2
|{\bf \Theta}\rangle_3.\nonumber
\end{eqnarray}
Next they swap the states of $A_1$ and $B_1$. This operation
utilizes two ebits and can be performed using two ordinary
teleportations in both directions. This yields
\begin{eqnarray}
\frac{1}{\sqrt{2}}
(|0\rangle_1^A|0\rangle_1^B+|1\rangle_1^A|1\rangle_1^B U_{23}^A
U_{23}^B) |{\bf \Psi}\rangle_2
|{\bf \Theta}\rangle_3,\nonumber\\
\frac{1}{\sqrt{2}} (|1\rangle_1^A|0\rangle_1^B
U_{23}^B+|0\rangle_1^A|1\rangle_1^B U_{23}^A) |{\bf \Psi}\rangle_2
|{\bf \Theta}\rangle_3.\nonumber
\end{eqnarray}
Next, they perform (\ref{localT}) again. The resulting states are
\begin{eqnarray}
|{\bf 0'}\rangle_1 |{\bf \Psi}\rangle_2 |{\bf \Theta}\rangle_3
\quad \mbox{and}\quad |{\bf 1'}\rangle_1 U_{23}^A U_{23}^B|{\bf
\Psi}\rangle_2 |{\bf \Theta}\rangle_3.\nonumber
\end{eqnarray}
Finally, they apply $H_A$ and $H_B$ again and obtain
\begin{eqnarray}
|{\bf 0}\rangle_1 |{\bf \Psi}\rangle_2 |{\bf \Theta}\rangle_3
\quad \mbox{and}\quad |{\bf 1}\rangle_1 U_{23}^A U_{23}^B|{\bf
\Psi}\rangle_2 |{\bf \Theta}\rangle_3.\nonumber
\end{eqnarray}

As we have already noted, two local C-NOT transformations are
equivalent to a nonlocal C-NOT transformation. Thus, from the last
expression it follows that a nonlocal C-NOT is applied on pairs 2
and 3 (with pair 2 as the control and pair 3 as the target) if and
only if the state of the first pair is $|{\bf 1}\rangle$. Thus
this protocol implements the nonlocal Toffoli gate and utilizes
two ebits, which are needed to swap two states in the intermediate
stage. Due to linearity of quantum mechanics all these arguments
will hold also in the case of arbitrary superposition of initial
states.

To conclude, from Eqs. (\ref{bound}, \ref{test}, \ref{ET}) we
obtain the following lower bound on the number $N_T$ of Toffoli
gates required to carry out a computation
\begin{equation}
N_T \geq \frac{|E_{in}^{test}-E_{out}^{test}|}{2}\ .
\end{equation}
We illustrate this result on the example of Shannon data
compression. We were led to consider this particular example by
our research in multi-particle entanglement compression
\cite{Tstatecompression}. In fact, this is how we discovered this
method in the first place.

The method of classical compression of $n$-bit source-string of
$0$'s and $1$'s, where $p$ is the probability of each bit to be
equal $1$, is based on the fact that the most probable (typical)
strings, generated by the source will contain $np$ ones when $n$
is large \cite{coverthomas}. If $p\neq {1 \over2}$ then the
Shannon entropy of the source $H(p)$ is smaller than 1 and the
number of typical strings, $2^{nH(p)}$, is less than the total
number of strings $2^n$. Thus, a message generated by the source
can be compressed to a shorter message.

We consider a "Shannon compressor" - a classical reversible
circuit which receives as input an $n$ bit string which contains
$np$ ones (i.e. a typical string) and outputs a compressed version
of the string in which only the first $\log \left( {n\atop
np}\right)\simeq n H(p)$ bits carry information and the other
$n(1-H(p))$ redundant bits are set to some standard sequence, e.g.
to all 0's:
\begin{equation}
x_1x_2...x_n \rightarrow f_1f_2...f_{nH}0_{nH+1}...0_n.
\end{equation}

Our goal is to find a lower bound on the number of Toffoli gates
needed to build the "Shannon compressor". We take the initial test
state to be the uniform superposition of states with $np$ ones:
\begin{eqnarray}\label{input}
|\Psi_{in}^{test}\rangle=N\sum_{x_i\in\{0,1\},\sum_i x_i = np}
|{\bf x_1}\rangle|{\bf x_2}\rangle...|{\bf
x_n}\rangle,~~~~~\nonumber
\end{eqnarray}
where $N= \left({n\atop np}\right)^{-1/2}$. The output state is:
\begin{eqnarray}
|\Psi_{out}^{test}\rangle=N \sum_{f_i \in \{0,1\}} |{\bf
f_1}\rangle|{\bf f_2}\rangle...|{\bf f_{nH}}\rangle |{\bf
0_{nH+1}}\rangle...|{\bf 0_n}\rangle.\nonumber
\end{eqnarray}

\begin{figure}
\epsfxsize=3.5truein \centerline{\epsffile{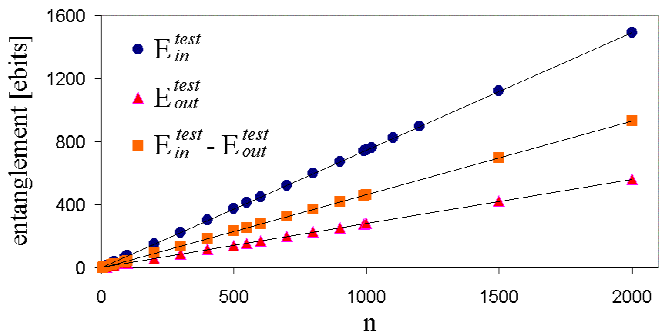}}
\caption[]{Entanglement as a function of string length $n$ for
$p=0.8$. The number of typical strings was calculated using non
approximated ${n \choose np}$. } \label{comp_ent}
\end{figure}

For fixed value of $p$ we can calculate the entanglement of
$|\Psi_{in}^{test}\rangle$ and $|\Psi_{out}^{test}\rangle$. The
entanglement $E_{out}^{test}$ is easy to calculate: it equals the
number of output redundant pairs, i.e. $E_{out}^{test}=n(1-H(p))$.
We have calculated $E_{in}^{test}$ using a combination of
analytical and numerical techniques which will be described in
\cite{Tstatecompression}. Fig. \ref{comp_ent} presents our results
for $p=0.8$. A linear dependance of $E_{in}^{test}-E_{out}^{test}$
on $n$ is obtained. Thus, the number of Toffoli gates needed to
perform Shannon compression grows at least linearly with $n$. For
instance for $p=0.8$ we need at least $0.2332n$ Toffoli gates.
Inspired by our numerical result, Buhrman has found, using a
completely different technique, an analytical proof of this lower
bound \cite{HB}.

In summary we have addressed the problem of evaluating the number
of Toffoli gates needed to perform classical reversible
computations. We have proposed a method based on quantum
information concepts in which strings of classical bits are mapped
into sequences of special nonlocal quantum states and classical
reversible computations are mapped onto unitary transformations of
these quantum states. The nonlocal properties of these
transformations provide information about the classical reversible
computation. In particular, if the unitary transformation is
nonlocal then the corresponding classical reversible circuit
cannot be built solely from one- and two-bit gates.  The amount of
non-locality possessed by the unitary transformation associated
with any classical reversible computation provides a lower bound
on the number of Toffoli gates needed for this computation.

As an example we considered classical Shannon compression and
calculated the amount of non-locality of the associated unitary
transformation. According to our numerical results, the lower
bound on the number of Toffoli gates grows linearly with the size
of the string $n$. Thus quantum methods can provide fundamental
insights about classical computation.

We hope that our approach may prove useful for other problems
concerning classical reversible computation.

\begin{acknowledgments}
We thank M. Ben-Or, H. Buhrman, N. Linden, M. Santha, U. Vazirani
and T. Short for useful comments. This work was supported by the
European IST-FET project RESQ, by the Communaut\'e Fran{\c{c}}aise
de Belgique grant No. ARC 00/05-251, by the IUAP program of the
Belgian government grant No. V-18.
\end{acknowledgments}

\end{document}